\documentstyle[aps,prb,epsf,floats]{revtex}

\begin{document}

\draft 

\twocolumn[{
\widetext

\title{
Identification of excitons in conjugated polymers: a density matrix renormalisation group study}

\author{
M. Boman$^\dagger$
}

\address{
School of Physics, University of New South Wales, Sydney, NSW 2052,
Australia \and Centre for Molecular Materials, The University of
Sheffield, S3 7RH Sheffield, United Kingdom.
}

\author{
R. J. Bursill
}

\address{
School of Physics, University of New South Wales, Sydney, NSW 2052,
Australia.
}

\maketitle

\mediumtext

\begin{abstract}

This work addresses the question of whether low-lying excitations in 
conjugated polymers are comprised of free charge-carriers or excitons. 
States are characterised as bound or unbound according to the scaling 
of the average particle-hole separation with system size. We critically 
examine other criteria commonly used to characterise states. The polymer 
is described by an extended Hubbard model with alternating transfer 
integrals. The model is solved by exact diagonalisation and the density 
matrix renormalisation group (DMRG) method. We demonstrate that the DMRG 
accurately determines excitation energies, transition dipole moments and 
particle-hole separations of a number of dipole forbidden ($A_g$) and 
dipole allowed ($B_u$) states.

Within a parameter regime considered reasonable for polymers such as 
polyacetylene, it is found that the charge gap, often used to define the 
exciton binding energy, is not a good criterion by which to decide 
whether a state is bound or unbound. The essential non-linear optical 
state $mA_g$ is found to mark the onset of unbound excitations in the 
$A_g$ symmetry sector. In the $B_u$ symmetry sector, on the other hand, 
it is found that {\it all} low lying states are unbound and that there 
is no well defined $nB_u$ state. That is, the $1B_u$ state marks the 
onset of unbound excitations in this sector.

\end{abstract}



}] \narrowtext

\section{Introduction} \label{Sec: Introduction}

The current interest in conjugated polymers lies to a large extent in 
their optical properties \cite{ICSM}. Conjugated polymers exhibit strong 
luminescence, and large and ultrafast nonlinear optical (NLO) response 
\cite{Etemad}. This has lead to technological opportunities and, for 
instance, polymer light-emitting diodes are now widely produced 
\cite{Burroughes}. These optical properties are associated with the 
delocalised $\pi$-electron system of the conjugated polymers and, in 
particular, the low-lying excitations. However, the nature of these 
excitations is not fully understood, and has been a subject for 
fundamental research on conjugated polymers in recent years \cite{ICSM}.

A central issue is whether low-lying excitations are comprised of 
free charge-carriers or excitons. If the Coulomb interaction between the 
oppositely charged particles and holes is strong, excitons are formed, 
i.e.\ bound particle-hole pairs, in which the motions of the particle 
and the hole are strongly correlated. On the other hand, if the Coulomb 
interaction is effectively screened, then the particles and holes are 
only very weakly bound and move essentially independently as free 
charge-carriers.

Different criteria have been used in the literature to discriminate 
between the states, leading to contradictory conclusions concerning the 
nature of the low-lying states. Some commonly used criteria are: the 
charge gap, $E_{\text g}$ \cite{Shuai Eb,Yaron,Campbell}; and the essential 
NLO states, $mA_{g}$ and $nB_{u}$ \cite{Guo,Chandross}.

The charge gap is defined as the sum of the energies for removing and 
adding an electron to the neutral system:
\begin{equation}
E_{\text g}=E(N+1)+E(N-1)-2E(N).
\end{equation}
where $E(N)$ refers to the ground state energy of the neutral system (where the band is half filled).
For the Hubbard model with only on-site Coulomb interaction, the charge 
gap coincides with the lowest optical excitation, i.e. the optical gap 
\cite{Shuai Eb}. This is, of course, also true for the tight-binding (or 
H\"{u}ckel) model, which is an independent-electron model, where the 
charge gap is the onset of the delocalised states in the conduction 
band. However, this is generally not true in the case of longer range 
Coulomb interactions since there are states below the charge gap. It has 
been argued that these states are exciton states since they, for 
instance, appear energetically within the tight-binding band gap. Thus, 
the charge gap has been used to discriminate between the states: states 
below the charge gap correspond to excitons and states above the charge 
gap correspond to free charge-carriers. Consequently, for a state of 
energy $E$, the  binding energy $E_{\text b}$ is defined as
\begin{equation}
E_{\text b} \equiv E_{\text g} - E.
\label{eq: binding energy}
\end{equation}
We believe that this criterion can be seriously criticised: (i) the 
charge-gap energy is in general not an eigenenergy of the system, but 
rather mixes ground state energies of three, all differently charged, 
systems; (ii) the criterion does not distinguish between different 
symmetry sectors; and (iii) the criterion does not directly measure the 
motion of particles and holes, but is based on total energies only. In 
this work, we have directly calculated the relative motion of particles 
and holes for different states in each symmetry sector in order to 
identify excitons.

It should be noted that the use of the Hartree-Fock bandgap in the 
literature \cite{Yaron} is merely an approximate way of calculating the 
charge gap. This follows directly from Koopmans' theorem, and we will 
therefore not treat the Hartree-Fock bandgap as separate criterion for 
excitons. It is well known that the Hartree-Fock bandgap systematically 
overestimates the true charge gap.

In works on NLO properties of conjugated polymers the most important 
channels for such processes have been identified
\cite{Heflin,Soos NLO,McWilliams,Dixit},
leading to a phenomenological model for third-order nonlinearity that is 
based on only the four most essential states \cite{Guo}. These states 
are the ground state $1A_{g}$, the lowest dipole allowed state $1B_{u}$, 
the $mA_{g}$ and the $nB_{u}$. The $mA_{g}$ is defined as the state 
which has the strongest dipole \cite{foot_dipole} coupling (or 
transition moment) to the $1B_{u}$, and the $nB_{u}$ is defined as the 
state which has the strongest dipole coupling to the $mA_{g}$, apart 
from the $1B_{u}$.

In addition, it was found that there is sudden increase in the
particle-hole separation at the $mA_{g}$ and the $nB_{u}$, i.e.\ all 
states below these states have more tightly bound particles and holes 
\cite{Guo}. It has therefore been argued that the $mA_{g}$ and the 
$nB_{u}$ are the lowest lying free charge-carrier states and useful 
criteria for the identification of excitons. The binding energy is 
defined analogously to (\ref{eq: binding energy}), but with the charge gap 
replaced with the $mA_g$ or $nB_u$ energy depending on which symmetry 
sector is of interest.

Although intriguing, it is not clear that states below the $mA_{g}$ and 
the $nB_{u}$ really correspond to excitons. The smaller particle-hole 
separation may simply be a consequence of system confinement. In 
principle one needs to study an infinite system to resolve this issue, 
whereas the results of \cite{Guo} pertain to an oligomer of $N=8$ carbon 
atoms. Another approach is to look at how the particle-hole separation 
scales with the system size $N$. In this work, we have calculated the 
particle-hole separation for different system sizes. The particle-hole 
separation scales differently for free charge-carriers and bound states.

In another study, the particle-hole separation was studied as a function 
of system size \cite{Yaron}, with electron correlation treated within 
the singly excited configuration-interaction (SCI) approximation. It was 
found that states of energy significantly below (above) the Hartree-Fock 
band gap were bound (unbound). Excitons are many body excitations and, 
for instance, it has been found that the exciton binding energy is 
sensitive to the inclusion of higher-order correlation through 
perturbation theory \cite{Suhai,Liegener}. It is therefore important to 
treat the electron correlation accurately when assessing exciton 
criteria. In this work, we have used exact diagonalisation of the 
Hamiltonian for systems ($N\leq 10$), and the Density Matrix 
Renormalisation Group method for longer systems ($N\leq 50$).

The methodology of this work is outlined in Sec.\
\ref{Sec: Methodology}, which contains definitions of calculated 
quantities and a description of the computational methods. The 
model Hamiltonian is defined in Sec.\ \ref{Sec: Model}, particle-hole 
separation definitions are given in Sec.\
\ref{Sec: ph sep def} where the basis of the scaling analysis is 
outlined, the ionicity is defined in Sec.\ \ref{Sec: Ionicity def} and 
the numerical solution of the model is described in Sec.\
\ref{Sec: computational methods}. A more detailed derivation of the 
scaling behaviour in the non-interacting limit is given in Appendix 
\ref{Huckel_appendix}. Results are presented and discussed in Sec.\ 
\ref{Sec: Results and discussion}. The accuracy of the DMRG method is 
demonstrated in Sec.\ \ref{Sec: Accuracy}, the energy spectra and 
transition moments are given in Sec.\ \ref{Sec: energy and trm}, results 
for the particle-hole separation follow in Sec.\ \ref{Sec: ph sep res}, 
whilst ionicity results are given in Sec.\ \ref{Sec: Ionicity res}. 
Finally, conclusions are given in Sec.\ \ref{Sec: Conclusions}.

\section{Methodology} \label{Sec: Methodology}

\subsection{Model} \label{Sec: Model}

As a generic model for conjugated polymers we study the extended 
Hubbard model with alternating hopping integrals and on-site and 
nearest-neighbour electron-electron interactions:
\begin{eqnarray}
\hat{H}
& \equiv &
-t\sum_{i,\sigma} [1-(-1)^{i}\delta]
(\hat{c}_{i\sigma}^{\dagger}\hat{c}_{i+1\sigma} + \mbox{ h.c.}) +
\nonumber \\
&   &
\frac{U}{2}\sum_{i}(\hat{n}_{i}-1)(\hat{n}_{i}-1)
+V\sum_{i}(\hat{n}_{i}-1)(\hat{n}_{i+1}-1),
\label{eq: hamiltonian}
\end{eqnarray}
where $\hat{c}_{i\sigma}$ annihilates a $\pi$-electron with spin 
$\sigma$ on site $i$ and $\hat{n}_{i}$ is the occupation number operator 
for site $i$. We have studied systems with an even number of sites $N$, 
and open boundary conditions. The Hamiltonian is a paradigm for 
conjugated polymers in which excitons can exist.

For convenience we take $t=1$, which sets the energy scale. Values for 
the dimerisation in the range 0.07--0.15 have been proposed for 
polyacetylene, polydiacetylene, and
poly({\it para}-phenylenevinylene) \cite{Soos delta,Shuai Eb}. We choose 
$\delta = 0.1$ as a typical value for conjugated polymers. Optical 
absorption data suggest a $U$ value of 2.25--2.75 \cite{Maz U}, whereas 
{\it ab initio} calculations suggest a slightly higher value of 3.6 
\cite{Konig}. We have taken $U=3$ as a reasonably realistic value.
For the nearest neighbour interaction we use the standard value $V=0.4U$ 
\cite{Guo,Shuai Eb}. Thus, the chosen values of the parameters are: 
$t=1$, $\delta = 0.1$, $U=3$, and $V=1.2$. These values are used 
throughout the paper, unless otherwise stated.

Besides conserving the total number of particles, the Hamiltonian possesses spatial symmetry,
$\hat{C}_{2h}\in \{A_{g},B_{u}\}$, charge conjugation symmetry, 
$\hat{J}\in \{+,-\}$, and spin symmetry, e.g.\
$\hat{S}\in \{0,1,2\ldots\}$ \cite{Cizek}. In this paper we are concerned with neutral, singlet states (with a half-filled band and
$\hat{S} = 0$), except for the charged states used in determining $E_{\text g}$. The singlet ground state is even under spatial inversion and charge conjugation, and is 
denoted $1^{1}A_{g}^{+}$, or simply $1A_g$. For this work we only need 
to consider the ground state symmetry sector and the sector to which it 
is dipole coupled. Since the dipole operator is odd under inversion and 
charge conjugation, states in the dipole coupled sector are denoted 
$j^{1}B_{u}^{-}$, or simply $jB_{u}$, where $j$ is the state number.

\subsection{Particle-hole separation} \label{Sec: ph sep def}

In this paper, we consider the relative motion of particles and holes as 
a direct way of identifying excitons. In particular, we calculate the 
average particle-hole separation. Excitons have small particle-hole 
separations which remain finite as the system size is increased. By 
contrast, the average separation between two free charge-carriers 
increases indefinitely with system size. Indeed, for a completely 
independent particle and hole, each with a uniform probability ($1/N$) 
of being on any given site, the leading term in the average separation 
is proportional to $N$.

In reference \cite{Guo} Guo {\it et al.} used the density-density
correlation function $C(i,j)$ as a signifier which distinguishes bound
and unbound states:
\begin{equation}
C(i,j) \equiv
\left\langle
( \hat{n}_i - \left\langle \hat{n}_i \right\rangle ) 
( \hat{n}_j - \left\langle \hat{n}_j \right\rangle )
\right\rangle.
\label{eq: charge corr}
\end{equation}
$C(i,j)$ correlates a charge fluctuation on site $i$ to a charge 
fluctuation on site $j$. A positive value means that an excess (deficit) 
of charge on site $i$ correlates with an excess (deficit) on site $j$. A 
negative value correlates an excess with a deficit. In the context of 
excitons, the concept of quasi-particles is a way of representing an 
aggregation of electrons which leads to an excess of charge in a region, 
which may extend over several sites. Likewise, a hole represents deficit 
of charge. Now, for sufficiently large $N$, $C(i,j)$ is negative 
(positive) for $|i-j|$ odd (even). We thus consider the odd distance 
contributions when defining the probability distribution from which to 
measure the separation of particles and holes. Other definitions, 
utilising both positive and negative contributions, or $|C(i,j)|$ are 
possible, and yield the same qualitative results.

In \cite{Guo} it was found that there is a noticeable change in the 
nature of the decay of $C(i,j)$ at the $mA_g$ and $nB_u$, the
particle-hole separation being considerably larger than for the lower 
lying states. However, calculations were restricted to $N=8$. We have 
used the DMRG method to study substantially larger systems (up to 
$N=50$). Most importantly, this makes it possible to discriminate 
between bound and unbound states from the {\it scaling} of $C(i,j)$ or 
the particle-hole separation with $N$. 

We define an averaged \cite{foot_averaging}, centered correlation 
function for a system of size $N$ with open boundary conditions as:
\begin{equation}
C_{N}(l) \equiv \left\{
  \begin{array}{ll}
    C(\frac{N-l}{2},\frac{N+l}{2}) & \mbox{$l$ even} \\
    \frac{1}{2}\left[C(\frac{N-l+1}{2},\frac{N+l+1}{2})+
    C(\frac{N-l-1}{2},\frac{N+l-1}{2}) \right] & \mbox{$l$ odd.} \\
  \end{array}
\right. .
\label{eq: finite corr}
\end{equation}
As discussed, we define the average particle-hole separation (in units 
of chemical bonds) by regarding the negative values of 
$C_{N}(l)$ as a probability distribution viz.
\begin{equation}
\left\langle l \right\rangle_{N} \equiv 
\frac{\sum_{l=0}^{N-1}l(|C_{N}(l)|-C_{N}(l))}
{\sum_{l=0}^{N-1}(|C_{N}(l)|-C_{N}(l))}.
\label{eq: ave ph sep}
\end{equation}

\begin{figure}[htbp]
\centerline{\epsfxsize=8.4cm \epsfbox{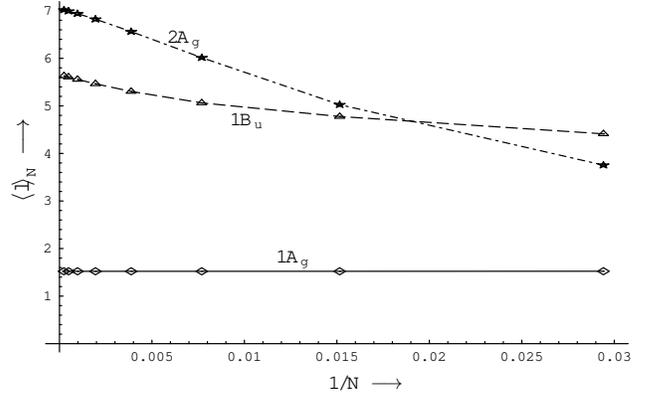}}
\caption{
Particle-hole separation $\left\langle l \right\rangle_N$ of the $1A_g$, $2A_g$ and $1B_u$ 
states in the non-interacting limit 
($U=V=0$) plotted as a function of $1/N$.
}
\label{Fig: ph sep scaling}
\end{figure}

It is instructive to consider the two extreme cases of very weak and 
very strong electron-electron interaction, since the identification of 
bound and unbound states is unambiguous in these cases. In the
non-interacting limit ($U=V=0$), we combine exact analytical results 
with exact diagonalisations of systems of up to 4098 sites in order to 
determine the scaling of the correlation function and particle-hole 
separation with $N$. Details are given in Appendix 
\ref{Huckel_appendix}. The main point is that, as can be seen 
from Fig.\ \ref{Fig: ph sep scaling}, the ground state ($1A_g$) and the
unbound states ($1B_u$ and $2A_g$) have particle-hole separations which  
scale very differently with $N$. For the ground state we have 
exponential convergence
\begin{equation}
\left\langle l \right\rangle_N = a
\left[ 1 + O\left(e^{-\alpha N}\right) \right],
\label{eq: bound scaling}
\end{equation}
and for the unbound excited states we have very slow convergence
\begin{equation}
\left\langle l \right\rangle_N = b
\left[ 1 + O\left( N^{-1} \right) \right].
\label{eq: unbound scaling}
\end{equation}

In the limit of strong interactions, ($U,V\gg t$), it was shown by Guo 
{\it et al.\ }\cite{Guo} that the states can be identified solely from 
their total energy. At $t=0$, the analysis is trivial: A particle 
(hole) is a doubly-occupied (empty) site. The ground state has zero 
energy, bound states occur at $U-V$, and unbound states at $U$. Since 
the Hamiltonian consists only of the $U$-$V$ potential, the bound states 
have a particle and a hole next to one another, the correlation function 
vanishes for $l>1$, and the particle-hole separation converges 
immediately with $N$, i.e.\
$\left\langle l \right\rangle_{N} \equiv 1$. For unbound states,
at energy $U$, it is easy to show that
$\left\langle l \right\rangle_{N}$ increases linearly 
with $N$, i.e.\ as for independent particles and holes.

In addition to the correlation function, we consider a {\it reduced 
correlation function}: the difference between the correlation function 
of an excitation, $C_{N}(l)$, and that of the ground state, 
$C_{N}^{\text{(GS)}}(l)$:
\begin{equation}
C_{N}^{\text{(red)}}(l) \equiv
N \left[ C_{N}(l) - C_{N}^{\text{(GS)}}(l) \right].
\label{eq: red corr}
\end{equation} 
The motivation for studying $C_{N}^{\text{(red)}}(l)$ is that it only 
measures {\it changes} in the charge fluctuations induced by an 
excitation. That is, the effect of creating particles and holes, whose 
motion we are interested in. By analogy with (\ref{eq: ave ph sep}), we 
also define a reduced particle-hole separation
$\left\langle l \right\rangle_{N}^{\text{(red)}}$.

\begin{figure}[htbp]
\centerline{\epsfxsize=8.4cm \epsfbox{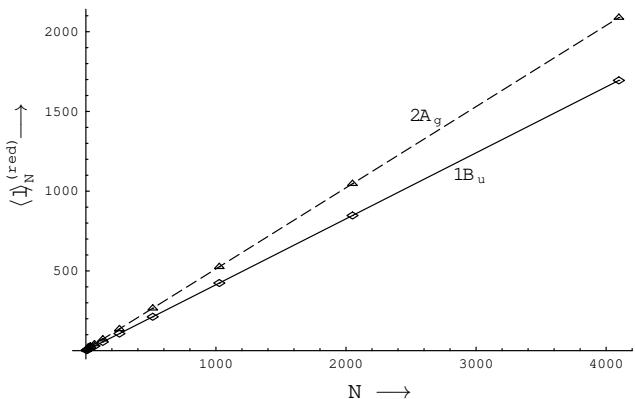}}
\caption{
Reduced particle-hole separation
$\left\langle l \right\rangle^{\text{(red)}}$ of the $1B_u$ and 
$2A_g$ states in the non-interacting limit ($U=V=0$) plotted as a 
function of $N$.
}
\label{Fig: red ph sep scaling}
\end{figure}

In the non-interacting case, the reduced particle-hole separation for
unbound states scales linearly with $N$, as shown in Fig.\
\ref{Fig: red ph sep scaling}. As mentioned above, this is how the 
separation of a completely independent particle and hole scale. In the 
strongly correlated limit, $C_{N}^{\text{(GS)}}(l)\equiv 0$ and so
$\left\langle l \right\rangle_{N}^{\text{(red)}}
\equiv
\left\langle l \right\rangle_{N}$
and hence
$\left\langle l \right\rangle^{\text{(red)}}_{N} \equiv 1$
for bound states and 
$\left\langle l \right\rangle^{\text{(red)}}_{N} \propto N$
for unbound states. That is, unbound 
states scale in the same way in both limits. This suggests that, for 
general interaction strength, the reduced particle-hole separation 
scales linearly for unbound states.  

In summary, for unbound states, the particle-hole separation
$\left\langle l \right\rangle_{N}$ converges slowly with $N$ and the reduced 
particle-hole separation
$\left\langle l \right\rangle_{N}^{\text{(red)}}$ diverges linearly. By 
contrast, for bound states, both
$\left\langle l \right\rangle_{N}$ and
$\left\langle l \right\rangle_{N}^{\text{(red)}}$ 
converge rapidly to a finite value.

\subsection{Ionicity} \label{Sec: Ionicity def}

In the limit of strong electron-electron interaction, $U\gg t$, the 
ground state is a pure spin-density-wave, i.e. it is solely a linear 
combination of covalent configurations (or Slater determinants), where 
there is exactly one electron on each site. It is therefore natural,
in this limit, to define a particle as a doubly occupied site, and a 
hole as an unoccupied site. Configurations which have particles and 
holes are referred to as ionic.
A general state can be written as 
$\Psi=\Psi^{\text{(covalent)}}+\Psi^{\text{(ionic)}}$.

Following \cite{McWilliams}, we define the ionicity as the  
number of particle-hole pairs \cite{foot_ionicity}. Thus, single 
particle-hole pair excitons are constructed from singly ionic 
configurations, biexcitons are constructed from doubly ionic 
configurations etc. The ionicity of a state is defined as the 
expectation value of the ionicity operator $\hat{I}$:
\begin{eqnarray}
\hat{I} & = & \frac{1}{2}\sum_{i} (\hat{n}_{i}-1)^{2}.
\label{eq: ionicity}
\end{eqnarray}

In addition to the ionicity, we have calculated the average ionic
particle-hole separation
$\left\langle l \right\rangle_{N}^{\text{(ionic)}}$ by considering only
the ionic part of the wavefunction. In contrast to
$\left\langle l \right\rangle_{N}$,
 $\left\langle l \right\rangle_{N}^{\text{(ionic)}}$ allows us to
probe changes within $\Psi^{\text{(ionic)}}$ alone, rather than
at the same time taking into account a change in the relative 
weights of $\Psi^{\text{(covalent)}}$ and $\Psi^{\text{(ionic)}}$.

\subsection{Computational methods } \label{Sec: computational methods}

Equation (\ref{eq: hamiltonian}) can be comfortably solved by
exact diagonalisation for systems of up to $N=12$ sites. For longer 
chains we turn to the Density Matrix Renormalisation Group (DMRG) method 
\cite{White}. The DMRG has been applied to (\ref{eq: hamiltonian}) in 
calculations of ground state and triplet energies \cite{pang}, 
charge densities \cite{pang}, the charge gap and $1B_u$ energies 
\cite{Shuai Eb,2Ag}, the $2A_g$ energy \cite{2Ag}, 
polarisabilities \cite{polar}, and oscillator strengths \cite{boman}.

In this work we apply an infinite lattice DMRG algorithm \cite{White} to 
find the charge gap and a number of $A_g$ and $B_u$ states of
(\ref{eq: hamiltonian}). In addition to excitation energies,
we find the transition dipole moments between the $A_g$ and $B_u$ states 
as well as the density-density correlation function
(\ref{eq: finite corr}) and hence the particle-hole separation
(\ref{eq: ave ph sep}) and reduced separation
$\left\langle l \right\rangle_N^{\text{(red)}}$. At 
each iteration the superblock consists of a system block, an environment 
block and two extra sites. The initial system and environment blocks 
consist of two sites. The system and environment blocks are increased by 
two sites at a time until a superblock of $N=50$ sites is reached. We 
retain $m=230$ density matrix eigenstates in the basis truncation 
procedure.

The system, environment and super block Hamiltonians and density matrices are block diagonalised using the total particle number $\hat{N}$ and total $z$-spin $\hat{S}^z$ operators.
States from different symmetry sectors are found by projecting trial 
states from the iterative, sparse matrix diagonalisation procedure into 
the correct symmetry sector by means of the operators $\hat{C}_{2h}$, 
$\hat{J}$ and the spin-parity operator $\hat{P}$ \cite{spin_parity}. Because the density matrix commutes with the block $\hat{J}$ and $\hat{P}$ operators, the superblock states 
calculated are {\it exact} eigenstates of $\hat{J}$ and 
$\hat{P}$ at all stages of the calculation. They are also exact 
eigenstates of $\hat{C}_{2h}$ by construction because the environment 
block is the reflection of the system block. We check the DMRG program 
by ensuring that it reproduces the results of the exact diagonalisation 
program for the first two iterations. The exact diagonalisation program 
itself reproduces the $N=8$ site results from \cite{Guo}.

\section{Results and discussion} \label{Sec: Results and discussion}

\subsection{Accuracy of the DMRG calculations}
\label{Sec: Accuracy}

We begin with the non-interacting limit ($U=V=0$), where we 
have performed separate exact diagonalisations for long chains in 
order to evaluate the DMRG accuracy. The non-interacting case is of 
particular interest since it gives a worst case accuracy. Errors are 
expected to be smaller in the interacting case where particles are more 
localised in position space, the DMRG becoming exact in the atomic 
($t=0$) limit. 

In the DMRG calculations the initial two systems ($N=6,10$) are treated exactly. The errors for these systems are merely a result of the limited precision of the sparse matrix diagonalisation algorithm (the accuracy could be increased by running the programs at higher precision). The DMRG truncation error sets in at $N=14$ where the relative error is $\sim 10^{-5}$. The error increases with system size and is
$\sim 10^{-4}$ for $N=30$--$50$.

In addition to excitation energies, it is important to look at quantities such as transition moments and particle-hole separations, which pertain to the wavefunction rather than the energy. The truncation error for transition moments 
is $\sim 10^{-3}$ for $N=30$--$50$, i.e.\ one order of magnitude larger than for excitation energies, but well within what is acceptable for our 
purposes. The particle-hole separations 
have similar errors to the transition moments.

The DMRG is a truncated basis method with systematically reducible 
error. In the interacting case it is important to test the accuracy of 
calculations by varying the single source of error, the truncation 
parameter $m$, and checking for convergence. We asses the error by 
running the program for a number of values of $m$.
Convergence results for a number of quantities are given in Table \ref{convergence} for the $N = 50$ site system. It is found that the ground state energy converges to 5 or 6 figures and gaps are resolved to within 0.01\% for the charge gap and $1B_u$ states, with slower convergence for the higher excitations where errors range up to one or two percent. The errors in transition dipole moments are larger but are still small enough to make a clear identification of the essential NLO states. Errors range from around 0.5\% for the $1A_g\rightarrow 1B_u$ transition to 1--2\% for the $mA_g\rightarrow 1B_u$ transition.
The errors in the particle-hole separations are found to range from 0.1\% for the $1A_g$ state to 1--3\% for the higher excitations. Convergence is sufficiently good to allow the clear classification of 
states as bound or unbound on the basis of the scaling of the
particle-hole separation.

\begin{table}[htbp]
\caption{
The ground state energy $(1A_g)$, the charge gap $E_{\text g}$, the $1B_u$ and $4A_g$ energy gaps, various transition moments
$\left\langle jB_{u} | \hat{\mu} | 1A_{g} \right\rangle$ and particle-hole separations $\left\langle l \right\rangle_{50}$ for the $N=50$ site system 
calculated using the DMRG for a number of truncation parameter values 
$m$. 
}
\begin{tabular}{ccccccccc}
    & \multicolumn{4}{c}{Energy}              &
\multicolumn{2}{c}{
$\left\langle jB_{u} | \hat{\mu} |mA_{g} \right\rangle$
}        &
\multicolumn{2}{c}{
$\left\langle l \right\rangle_{50}$
}       \\
\cline{2-5} \cline{6-7} \cline{8-9} \\
\multicolumn{1}{c}{$m$} & \multicolumn{1}{c}{$1A_g$}       & 
\multicolumn{1}{c}{$E_{\text g}$}  & \multicolumn{1}{c}{$1B_u$} & 
\multicolumn{1}{c}{$4A_g$} & \multicolumn{1}{c}{$1B_u$}& 
\multicolumn{1}{c}{$3B_u$}& \multicolumn{1}{c}{$2B_u$}&
\multicolumn{1}{c}{$4B_u$}
\\ \hline
 64 & $-$102.97649 & 1.2190 & 1.1122 & 1.1761 & 1.479 & 9.472 & 1.883 & 1.942 \\
100 & $-$102.97925 & 1.2212 & 1.1058 & 1.5455 & 7.151 & 6.599 & 2.018 & 1.726 \\
150 & $-$102.97969 & 1.2213 & 1.1044 & 1.4801 & 7.065 & 6.847 & 2.061 & 1.632 \\
185 & $-$102.97996 & 1.2218 & 1.1026 & 1.3855 & 6.962 & 7.297 & 2.078 & 1.602 \\
230 & $-$102.98002 & 1.2218 & 1.1025 & 1.3431 & 6.945 & 7.330 & 2.085 & 1.596 \\
\end{tabular}
\label{convergence}
\end{table}
 
\subsection{Excitation energies and transition moments}
\label{Sec: energy and trm}

\begin{figure}[htbp]
\centerline{\epsfxsize=8.4cm \epsfbox{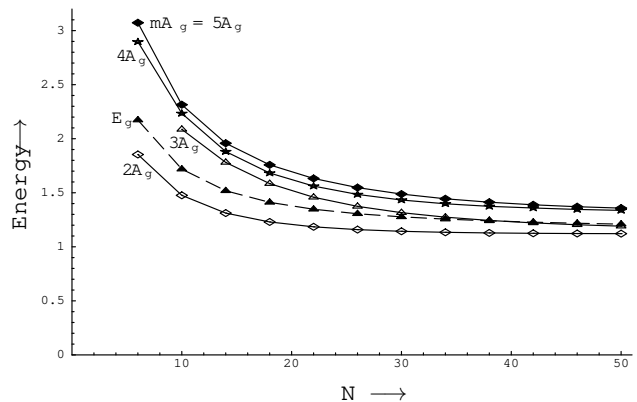}}
\caption{
Excitation energies (relative to the $1A_g$) in the $A_g$ symmetry 
sector as a function of $N$. The charge gap (dashed line) is shown
for comparison.
}
\label{Fig: Ag spectrum}
\end{figure}

\begin{figure}[htbp]
\centerline{\epsfxsize=8.4cm \epsfbox{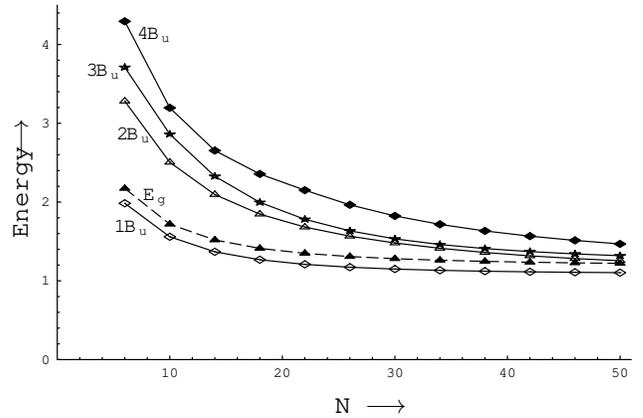}}
\caption{
Excitation energies (relative to the $1A_g$) in the $B_u$ symmetry 
sector as a function of $N$. The charge gap (dashed line) and is 
shown for comparison.
}
\label{Fig: Bu spectrum}
\end{figure}

The evolution of the energy spectra with system size is shown in 
Fig.\ \ref{Fig: Ag spectrum} and Fig.\ \ref{Fig: Bu spectrum} for the 
$A_g$ and $B_u$ sectors respectively. The charge gap is also shown for 
comparison. There is exactly one excitation below the charge gap for all
system sizes in the $B_u$ symmetry sector, and for most systems
in the $A_g$ sector. The $3A_g$ drops below the charge gap for
$N=42$--$50$.

Transition moments between various $A_g$ states and the $1B_u$ are given 
in Table \ref{Tab: jAg trm}. The $mA_g$ can be identified as the $5A_g$ 
for $N=10$--$50$. The transition moment with the $2A_g$ increases with 
system size, while transition moments with the $3A_g$ and the $4A_g$ 
decrease for $N\geq 22$. The charge gap is always below the $mA_g$ 
energy so the charge gap criterion will give lower exciton binding 
energies than the $mA_g$ criterion.

\begin{table}[htbp]
\caption{
Transition moments
$\left\langle jA_{g} | \hat{\mu} |1B_{u} \right\rangle$ for various states $jA_{g}$ and system sizes $N$.
}
\begin{tabular}{cccccc}
&\multicolumn{5}{c}{$\left\langle jA_{g} | \hat{\mu} |1B_{u} 
                                         \right\rangle$ } \\
\cline{2-6}
 $N$& $1A_g$ & $2A_g$ & $3A_g$ & $4A_g$ & $5A_g$ \\ \hline
  6 &  1.529 &  0.239 &  0.185 &  2.697 &  0.195 \\
 10 &  2.144 &  0.664 &  0.000 &  0.970 &  4.006 \\
 14 &  2.613 &  1.210 &  0.179 &  2.089 &  4.879 \\
 18 &  2.987 &  1.794 &  0.270 &  2.592 &  5.691 \\
 22 &  3.300 &  2.369 &  0.241 &  2.779 &  6.384 \\
 26 &  3.548 &  2.924 &  0.173 &  2.770 &  6.920 \\
 30 &  3.789 &  3.408 &  0.112 &  2.639 &  7.290 \\
 34 &  4.071 &  3.706 &  0.069 &  2.421 &  7.464 \\
 38 &  4.303 &  3.989 &  0.046 &  2.162 &  7.543 \\
 42 &  4.528 &  4.203 &  0.043 &  1.526 &  7.627 \\
 46 &  4.636 &  4.703 &  0.046 &  1.786 &  7.594 \\
 50 &  4.831 &  4.899 &  0.058 &  1.603 &  7.521 \\
\end{tabular}
\label{Tab: jAg trm}
\end{table}

The transition moments for the identification of the $nB_u$ are given
in Table \ref{Tab: jBu trm}. There are two qualitative differences
between these transition moments and those used for the identification
of the $mA_g$ (Table \ref{Tab: jAg trm}). Firstly, while the state
number of the $mA_g$ is basically constant ($m=5$, apart from $N=6$),  
the state number of the $nB_u$ changes with the system size
when the definition of Sec.\ \ref{Sec: Introduction} is
applied. It is the $5B_u$ at N=6, the $4B_u$ at $N=10,14$ and the
$2B_u$ at $N=18,22,26$.

\begin{table}[htbp]
\caption{
Transition moments 
$\left\langle jB_{u} | \hat{\mu} |mA_{g} \right\rangle$ 
for various states $jB_{u}$ and system sizes $N$.
}
\begin{tabular}{ccccccc}
&\multicolumn{6}{c}{$\left\langle jB_{u} | \hat{\mu} |mA_{g}
                                         \right\rangle$ } \\
\cline{2-7}
 $N$& $1B_u$ & $2B_u$ & $3B_u$ & $4B_u$ & $5B_u$ & $6B_u$ \\
\hline
  6 &  2.697 &  0.908 &  0.281 &  1.916 &  2.135 &  0.041 \\
 10 &  4.006 &  2.059 &  0.391 &  3.977 &  0.235 &  1.735 \\
 14 &  4.879 &  3.524 &  0.016 &  4.640 &  0.287 &  2.446 \\
 18 &  5.691 &  5.278 &  0.845 &  4.771 &  3.299 &  1.291 \\
 22 &  6.384 &  6.892 &  2.664 &  0.396 &  5.997 &  2.207 \\
 26 &  6.920 &  6.860 &  6.585 &  1.822 &  0.987 &  6.361 \\
%
\end{tabular}
\label{Tab: jBu trm}
\end{table}

Secondly, for many system sizes, the $nB_u$ is ill-defined in the
sense that there are several transition moments close to the maximum
value and to that of the $nB_u$. For instance, at $N=18$, the $1B_u$, 
$2B_u$ and $4B_u$ have very similar transition moments; at $N=22$, the 
$1B_u$, $2B_u$ and $5B_u$; and at $N=26$, the $1B_u$, $2B_u$, $3B_u$, 
and $6B_u$. It was found in \cite{Guo} that conduction-band states had 
strong dipole coupling to {\it several} neighbouring conduction-band 
states in the dipole coupled symmetry sector. This was seen to 
distinguish band-to-band transitions from transitions involving 
excitons. If the $mA_g$ is the onset of the conduction-band, i.e.\ the 
lowest lying unbound $A_g$ state, then our results imply that the $1B_u$ 
is the onset of the conduction-band in the $B_u$ symmetry sector. A 
stronger argument, based on the scaling of the particle-hole separation, 
is given in Sec.\ \ref{Sec: ph sep res}.

\subsection{Particle-hole separation}
\label{Sec: ph sep res}

\begin{figure}[htbp]
\centerline{\epsfxsize=8.4cm \epsfbox{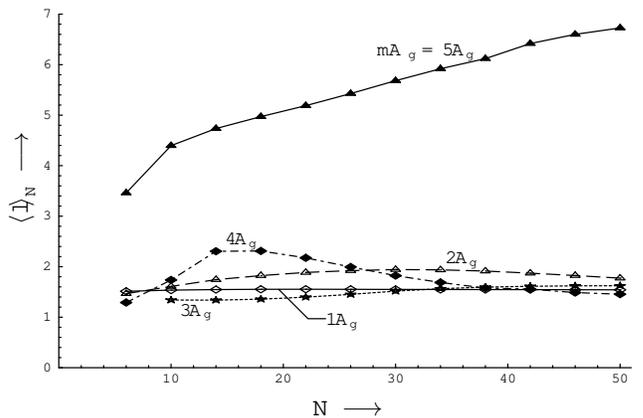}}
\caption{
Particle-hole separation $\left\langle l \right\rangle_N$ of the $A_g$ states as a function of 
$N$. The plot shows all states up to the $mA_g$.
}
\label{Fig: Ag norm ph sep}
\end{figure}

The particle-hole separation
$\left\langle l \right\rangle_N$ of the $A_g$ states is plotted as 
a function of $N$ in Fig.\ \ref{Fig: Ag norm ph sep}. We see that the 
$mA_g$ displays completely different behaviour from all the lower lying 
$A_g$ states. For the lower states, the average separation converges 
rapidly to the ground state (bulk limit) value of around 1.3 chemical 
bonds \cite{foot_minimum}. By contrast, for the $mA_g$ the separation is 
significantly larger for all $N$, exhibiting a far stronger $N$ 
dependence, and converging to a different limit from the bulk limit if 
converging at all. It follows that the $A_g$ states below the $mA_g$ are 
bound and the $mA_g$ is unbound.

\begin{figure}[htbp]
\centerline{\epsfxsize=8.4cm \epsfbox{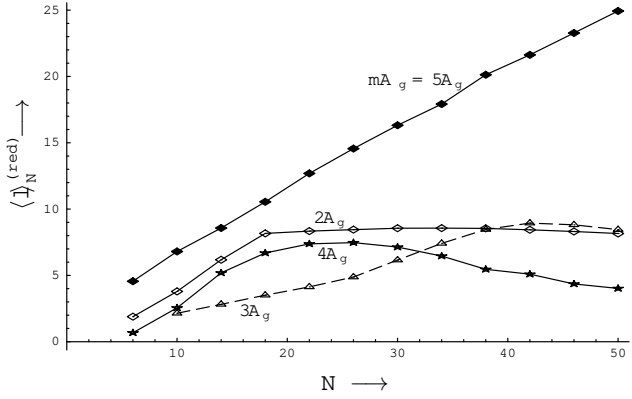}}
\caption{
The reduced particle-hole separation $\left\langle l \right\rangle^{\text{(red)}}_N$ of the $A_g$ 
states as a function of $N$. The plot shows all states up to the $mA_g$.
}
\label{Fig: Ag red ph sep}
\end{figure}

The striking difference between the $mA_g$ and the lower $A_g$ states is 
also borne out in the behaviour of the reduced particle-hole separation 
$\left\langle l \right\rangle^{\text{(red)}}_N$
depicted in Fig.\ \ref{Fig: Ag red ph sep}. The 
reduced separation of the $mA_g$ scales linearly with $N$, i.e.\ as for 
completely independent particles and holes. For $N=50$, the average 
separation is around 25 chemical bonds. The lower states, on the other 
hand, all have bounded separation. For instance, for the $2A_g$ 
$\left\langle l \right\rangle^{\text{(red)}}_N$ scales linearly up to $N=18$ where it levels off 
at around 8 chemical bonds. The interpretation of this is clear: the 
$2A_g$ is comprised of excitons with an average size of 8 bonds. When 
the system reaches a sufficiently large size the separation settles at 
the intrinsic average size of the exciton. For smaller systems, around 
or below the exciton size, the average separation is dictated by the 
system confinement.

In summary, the $A_g$ states below the $mA_g$ are bound, and the $mA_g$ 
is the lowest lying free charge-carrier state in this symmetry sector. 
We have only performed calculations for up to $N=50$, and thus in 
principle we cannot be certain that the reduced separation of the $mA_g$ 
does not eventually taper off. However, it is clear from our
calculations that the $mA_g$ behaves as a free charge-carrier state
for systems of up to 50 sites. If an upper bound does exist, then the
$mA_g$ must be extremely weakly bound. The $mA_g$ energy is therefore
a good reference energy when calculating binding energies in this
symmetry sector. The charge gap, on the other hand, falls amongst the
bound states. Some of the bound states are above the charge gap, while
others are below. Thus, the charge gap fails to discriminate between
bound and unbound states.

\begin{figure}[htbp]
\centerline{\epsfxsize=8.4cm \epsfbox{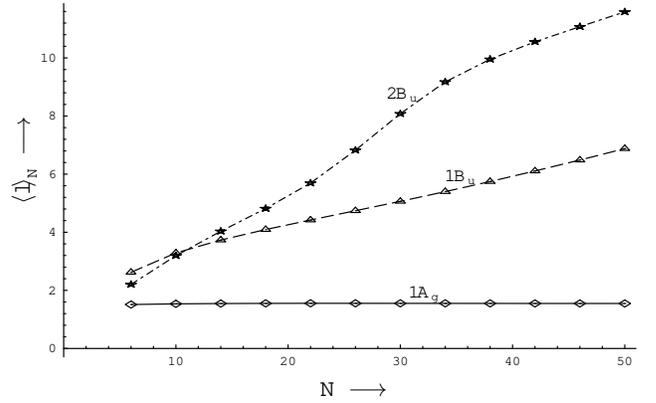}}
\caption{
Particle-hole separation $\left\langle l \right\rangle_N$ of the $B_u$ states as a function of 
$N$. The $1A_g$ state is plotted for reference.
}
\label{Fig: Bu norm ph sep}
\end{figure}

\begin{figure}[htbp]
\centerline{\epsfxsize=8.4cm \epsfbox{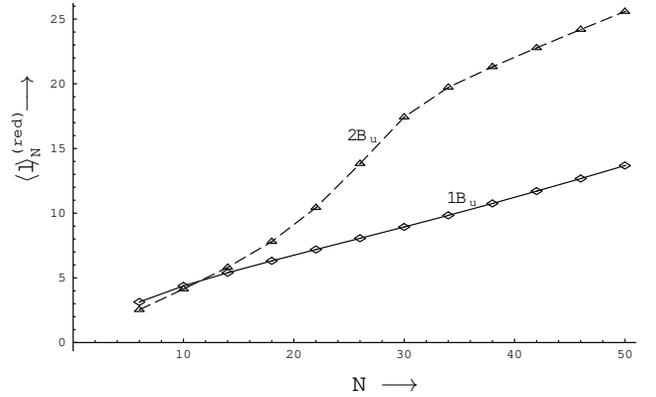}}
\caption{
The reduced particle-hole separation $\left\langle l \right\rangle^{\text{(red)}}_N$ of the $B_u$ 
states as a function of $N$.
}
\label{Fig: Bu red ph sep}
\end{figure}

We now turn to the dipole-allowed ($B_u$) symmetry sector. The
particle-hole separation is shown in Fig.\ \ref{Fig: Bu norm ph sep}. 
The separations depend strongly on $N$, indicating that there are no 
bound states in this sector. This is confirmed by the reduced
particle-hole separation, shown in Fig.\ \ref{Fig: Bu red ph sep}: The 
$1B_u$ scales almost perfectly linearly with $N$ and the $2B_u$ 
increases at least as rapidly over the plotted range. This explains the 
absence of a well defined $nB_u$ state above the $1B_u$ 
(see Sec.\ \ref{Sec: energy and trm}). That is, the $1B_u$ {\it 
is} the lowest unbound state in the $B_u$ sector. We note that the 
charge gap lies above the $1B_u$, incorrectly suggesting that the $1B_u$ 
is a bound state (see Fig.\ \ref{Fig: Bu spectrum}). As was the case
for the $A_g$ sector, the charge gap is not a useful criterion for 
determining whether states are bound and unbound in this symmetry 
sector.

\subsection{Ionicity} \label{Sec: Ionicity res}

\begin{figure}[htbp]
\centerline{\epsfxsize=8.4cm \epsfbox{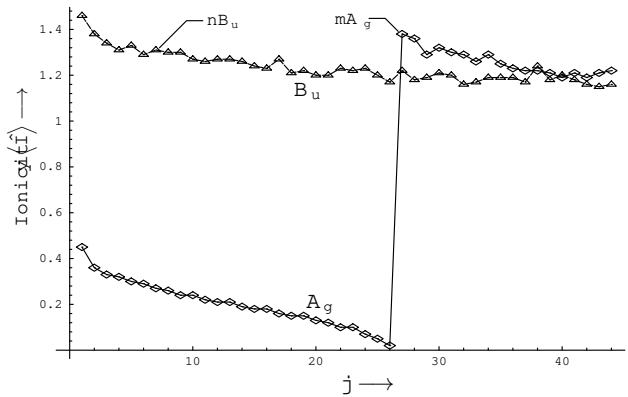}}
\caption{
The ionicity of the $jA_g$ and $jB_u$ states for $U=10$, $V=4$, $N=10$.
}
\label{Fig: mod Ag n_eff}
\end{figure}

The ionicity gives additional insight into the nature of the low-lying 
states. It is instructive to first study a case of relatively strong 
Coulomb interaction ($U=10$, $V=4$). As seen in
Fig.\ \ref{Fig: mod Ag n_eff}, there is a sudden jump in the ionicity
at the $mA_g$, for which
$\left\langle\hat{I}\right\rangle=1.4$. The ground state has an 
ionicity of 0.45, and an excitation to the $mA_g$ creates almost exactly 
one particle-hole pair. Lower lying states, on the other hand, follow a 
more continuous evolution with similar ionicity to that of the ground 
state. In fact, the ionicity decreases, the $(m-1)A_g$ being almost 
purely covalent. By contrast, there is no such jump in the $B_u$ 
symmetry sector at the $nB_u$. Instead all states have about the same 
ionicity, similar to that of the $mA_g$.

\begin{figure}[htbp]
\centerline{\epsfxsize=8.4cm \epsfbox{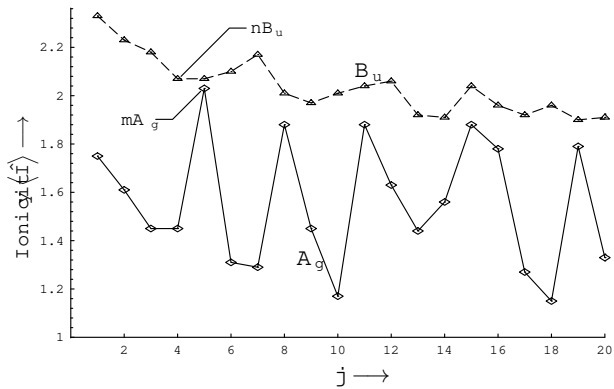}}
\caption{
The ionicity of the $jA_g$ and $jB_u$ states for $U=3$, $V=1.2$, $N=10$.
}
\label{Fig: Sh Ag n_eff}
\end{figure}

For the weaker, more realistic, interaction (see 
Sec.\ \ref{Sec: Model}), the ionicity is shown in 
Fig.\ \ref{Fig: Sh Ag n_eff}. Although there are large quantitative 
differences, many qualitative features remain. In the $A_g$ sector, the 
ionicity decreases continuously up to the $mA_g$ where a distinct jump 
occurs. Again, the ionicity of the $B_u$ states remains constant at 
around the same value as the $mA_g$. There is, however, one qualitative 
change, namely that the spectrum above the $mA_g$ is now a mixture of 
states with high and low ionicity. A picture where the $mA_g$ is the 
onset of a band of states which are all unbound appears to be too 
simplistic, at least from the $N=10$ site data.

The average ionic separation, shown in Table \ref{Tab: ionic sep}, also
shows a jump at the $mA_g$, from slightly more than 2 for the lower 
states to 3.5. In the other symmetry sector, the maximum ionic 
separation is 3.8 for the $nB_u$. However, the jump is less pronounced 
since the lower states all have relatively large separations of around 
3.0.

\begin{table}[htb]
\caption{
The average ionic separation of $jA_g$ and $jB_u$ states for $N=10$.
}
\begin{tabular}{ccc}
& \multicolumn{2}{c}{$\left\langle l \right\rangle_{10}^{\text{(ionic)}}$} \\
\cline{2-3}
j & $A_g$ & $B_u$ \\
\hline
1 & 2.23 & 2.97 \\
2 & 2.21 & 3.21 \\
3 & 2.04 & 2.82 \\
$4=n$ & 2.11 & 3.77 \\ 
$5=m$ & 3.54 & 3.08 \\
\end{tabular}
\label{Tab: ionic sep}
\end{table}

To conclude, the $mA_g$ excitation involves the creation of more 
particle-hole pairs than the lower lying states in the $A_g$ sector. 
Furthermore, the particle-hole pairs as less strongly bound. The large 
charge separation of the $mA_g$ is a result of two factors: a 
redistribution of weights in the wavefunction, from covalent to ionic 
configurations, and an increased charge separation within the ionic part 
of the wave function. By contrast, in the $B_u$ symmetry sector the 
difference between the $nB_u$ and the lower states is far less 
pronounced, a result which is consistent with the findings of 
Sec.\ \ref{Sec: ph sep res}.

\section{Conclusions} \label{Sec: Conclusions}

In this paper we have studied the particle-hole separation in the 
fundamental model of conjugated polymers---the dimerised, extended  
Hubbard model---in order to critically assess criteria commonly used to 
determine whether excitations are excitonic or consist of uncorrelated 
particle-hole pairs. The chosen parameter values are typical values used 
to describe polymers such as polyacetylene within the model. The model 
was solved for the charge gap and a number of states in the ground state 
($A_g$) and dipole allowed ($B_u$) symmetry sectors using the Density 
Matrix Renormalisation Group (DMRG) method for systems of up to $N=50$ 
sites. It was shown that the DMRG can be used to accurately compute 
energy gaps, transition dipole moments and particle-hole separations for 
these states, with relative errors ranging from a fraction of a percent 
for the lowest states to a few percent for higher states. The
particle-hole separations of bound and unbound states were shown to 
scale differently with system size, a fact which can be used to 
discriminate between the states.

It was found that the charge gap, often used to define the binding 
energy of excitons, is not a useful criterion by which to decide whether 
a state is bound or unbound. The results for the scaling of the
particle-hole separation show that there is an unbound state below the 
charge gap in the $B_u$ symmetry sector, and bound states states above 
the charge gap in the $A_g$ sector. Thus, it is possible to have states 
below the charge gap which are unbound, and states above the charge gap 
which are bound. In fact, the results were in keeping with the picture 
that the essential non-linear optical state $mA_g$ marks the onset of 
unbound states in the $A_g$ symmetry sector. On the other hand, in the 
$B_u$ sector it was found that all states were unbound for the 
parameters considered, and that there was no well defined $nB_u$ state. 
That is, the $1B_u$ state marks the onset of unbound excitations in the 
$B_u$ sector.

Finally, we note that, although the dimerised, extended Hubbard model is a fundamental model which is believed to provide a qualitative description of polymers such as polyacetylene, one needs to investigate the effects of long range Coulomb interactions, interchain coupling and electron-phonon effects before one can begin to make reliable statements about the nature of excitons in specific, real conjugated systems. We are currently developing efficient, vectorised DMRG codes with which we will be able to study these effects.

\acknowledgements

M. B. acknowledges the support of the Swedish Research Council for
Engineering Sciences (TFR), and R. J. B. acknowledges the support of the 
Australian Research Council. We would like to thank Dr W. Barford for 
extensive discussions throughout the course of this work and Dr B. T. 
Pickup from the University of Sheffield and Dr R. McKenzie and Dr C. 
Hamer from the University of New South Wales (UNSW) for useful 
discussions. M. B. also acknowledges positions as Honorary Visiting 
Fellow at UNSW and as Visiting Scholar at the Centre for Molecular 
Materials, Sheffield. We would like to thank Dr R. Standish and the New 
South Wales Centre for Parallel Computing for the use of the SGI 
Power Challenge facility on which the DMRG calculations were performed.

\appendix \label{appendix}

\section{Scaling of the correlation function and particle-hole
separation in the non-interacting limit}

\label{Huckel_appendix}

In this appendix we present some results for the correlation functions 
of excited states of (\ref{eq: hamiltonian}) in the non-interacting limit
($U=V=0$). In arriving at these results, we make use of well 
known analytical results for open and periodic systems and 
diagonalisations of open systems of up to 4098 sites.

The bulk correlation function is given by
\begin{eqnarray}
C_\infty(l)
& \equiv &
\lim_{N\rightarrow\infty} C_N(l)
\label{Eq: bulk corr}
\\
& = &
\left\{
       \begin{array}{ll}
       1/2 & l = 0 \\
       0   & l > 0\mbox{, even} \\
       - |F(l,\delta,+)|^2 - |F(l,\delta,-)|^2 & l\mbox{ odd,}
       \end{array}
\right.
\end{eqnarray}
where
\begin{equation}
F(l,\delta,\pm) \equiv \frac{1}{\pi}
\int_{-\pi/2}^{\pi/2} e^{il\theta}
\sqrt{
      \frac{
            \cos\theta \pm i \delta \sin\theta
           }{
            \cos\theta \mp i \delta \sin\theta
           }
     }
\, d\theta
\end{equation}
For $0 < \delta < 1$, the model has a gap $\Delta = 4 \delta t$ and 
exponentially decaying correlations
\begin{equation}
C_\infty(l) \sim e^{-l/\xi} \mbox{ as } l \rightarrow \infty
                                        \mbox{\space ($l$ odd),}
\label{Cinf_scaling}
\end{equation}
with correlation length
\begin{equation}
\xi = \frac{ 1 } { 2 \tanh^{-1} \delta }.
\end{equation}

For the ground state of a system of size $N$ we have
\begin{equation}
C_N^{\text{(GS)}}(l) \equiv C_\infty(l) \equiv 0 \mbox{ for all $l$ even,}
\label{CGS_l_even}
\end{equation}
and for odd $l$, $C_N^{\text{(GS)}}(l)$ approaches $C_\infty(l)$ 
exponentially fast viz
\begin{equation}
C_N^{\text{(GS)}}(l) = C_\infty(l) \left[
                                      1 + O\left( e^{-\alpha N} \right)
                               \right]\mbox{ ($l$ odd),}
\label{CGS_l_odd}
\end{equation}
for some positive constant $\alpha$ which is uniform over $l$. It 
follows that the ground state particle-hole separation converges 
rapidly, according to (\ref{eq: bound scaling}), where
\begin{equation}
a \equiv \lim_{m\rightarrow\infty}
\frac{
     \sum_{l = 1}^{m} (2 l - 1) \left| C_\infty(2 l - 1) \right|
    }{
     \sum_{l = 1}^{m} \left| C_\infty(2 l - 1) \right|
     }
\label{lave_inf}
\end{equation}
is the particle-hole separation defined from the bulk correlation 
function (\ref{Eq: bulk corr}). This can be seen in 
Fig.\ \ref{Fig: ph sep scaling} where 
$\left\langle l \right\rangle_N$ is plotted for the $1A_g$ state.

\begin{figure}[htbp]
\centerline{\epsfxsize=8.4cm \epsfbox{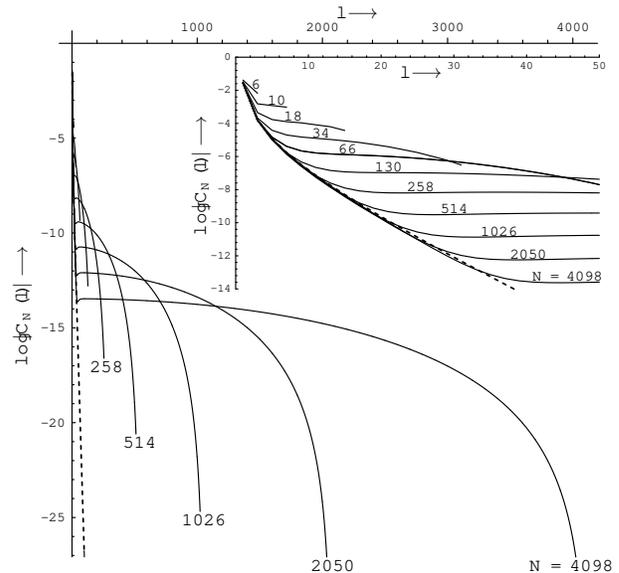}}
\caption{
$\log |C_N(l)|$ for the $1B_u$ state as a function of $l$ for $N=$ 6, 
10, 18, 34, 66, 130, 258, 514, 1026, 2050 and 4098 in the
non-interacting case ($U=V=0$). Also plotted is
$\log |C_\infty(l)|$, the bulk (ground state) value (dashed line). Only 
the odd values of $l$ are used. The inset shows the same plot on a 
smaller horizontal scale.
}
\label{Fig: logC}
\end{figure}

\begin{figure}[htbp]
\centerline{\epsfxsize=8.4cm \epsfbox{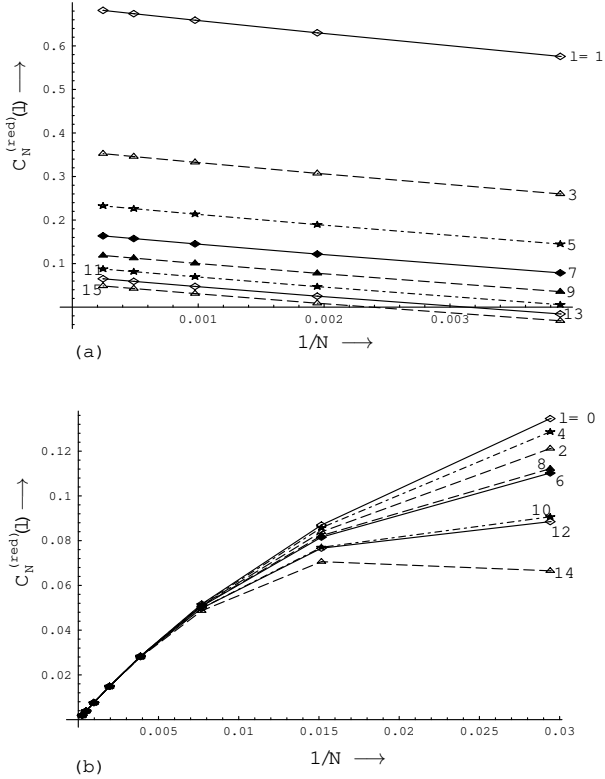}}
\caption{
Reduced correlation function $C_N^{\text{(red)}}(l)$ for the $1B_u$ state 
as a function of $1/N$ for various (a) odd and (b) even values of $l$ in 
the non-interacting case ($U=V=0$).
}
\label{Fig: reducedC}
\end{figure}

For the excited states the finite lattice correlation function $C_N(l)$ 
approaches the infinite lattice value $C_\infty(l)$ as 
$N\rightarrow\infty$ for any fixed $l$ \cite{foot_excitation}. However, 
as can be seen from Fig.\ \ref{Fig: logC}, where we plot $\log |C_N(l)|$ 
for the $1B_u$ states as a function of $l$ for various values of $N$, 
the convergence is very slow. This is shown in Fig.\
\ref{Fig: reducedC}, where we plot the reduced correlation function 
$C_N^{\text{(red)}}(l)$, defined by (\ref{eq: red corr}), as a function of $1/N$ for a number of values of $l$. We see clearly that the scaling behaviour of the excited state correlation function is
\begin{equation}
C_N(l) = C_\infty(l) + \frac{ C^{\text{(red)}}_\infty(l) }{ N }
         + \frac{ g(l) }{ N^2 } + \ldots,
\label{eq: scaling form}
\end{equation}
where $C^{\text{(red)}}_\infty(l)$ is the limiting, reduced correlation 
function viz
\begin{equation}
C^{\text{(red)}}_\infty(l) \equiv
           \lim_{N\rightarrow\infty} C^{\text{(red)}}_N(l).
\label{eq: C_in_red}
\end{equation}

\begin{figure}[htbp]
\centerline{\epsfxsize=8.4cm \epsfbox{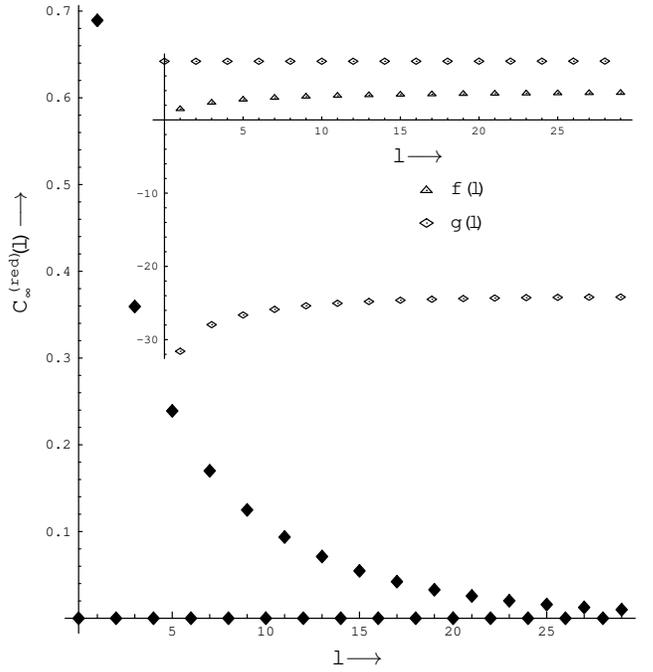}}
\caption{
Limiting reduced correlation function $C_\infty^{\text{(red)}}(l)$ for the 
$1B_u$ state as a function of $l$ in the non-interacting case ($U=V=0$).
The inset shows the scaling functions $f(l)$ and $g(l)$ ($f(l)$ is only 
defined for odd values of $l$ because $C_\infty^{\text{(red)}}(l)=0$ for 
even $l$).
}
\label{Fig: limreducedC}
\end{figure}

Using a polynomial fit of the form (\ref{eq: scaling form}) and 
extrapolating to the $N=\infty$ limit, $C^{\text{(red)}}_\infty(l)$ is 
calculated and plotted in Fig.\ \ref{Fig: limreducedC}. We 
see that $C^{\text{(red)}}_\infty(l)$ decays exponentially with $l$, 
and vanishes for even $l$. In fact, the scaling form
(\ref{eq: scaling form}) can be derived explicitly for the case of periodic boundary 
conditions, where it can be shown that
\begin{equation}
C^{\text{(red)}}_\infty(l) = \sqrt{ C_\infty(l) } f(l)
\label{eq: C_red_inf_decay}
\end{equation}
where the functions $f(l)$ and $g(l)$ rapidly approach non-zero 
constants as $l$ is increased (i.e.\ $C^{\text{(red)}}_\infty(l)$ decays 
exponentially and has correlation length $2\xi$). The functions $f$ and 
$g$, derived from the polynomial fit to (\ref{eq: scaling form}), are 
plotted in the inset of Fig.\ \ref{Fig: limreducedC}.

Now, the straightness of the curves in Fig.\ \ref{Fig: reducedC} 
indicates that the higher order terms in (\ref{eq: scaling form}) are 
small, and Fig.\ \ref{Fig: logC} indicates that (\ref{eq: scaling form}) 
holds for a range of $l$ values that is of $O(N)$. These results, 
together with the definition (\ref{eq: ave ph sep}), imply that the 
particle-hole separation for an excited state scales according to 
(\ref{eq: unbound scaling}), with limiting value
$b \equiv \lim_{N\rightarrow\infty}
\left\langle l \right\rangle_N$.
This can be 
seen from Fig.\ \ref{Fig: ph sep scaling}, where $\left\langle l \right\rangle_N$ is 
plotted as a function of $1/N$ for the $1B_u$ and $2A_g$ states. We note 
that $b \ne a$ i.e.\ the ground and excited states have 
different particle-hole separations in the bulk limit, even though their 
correlation functions approach the same limiting function. This is due 
to the fact that the third term in (\ref{eq: scaling form}) does 
not decay with $l$.

Finally, we note that, from (\ref{eq: red corr}), (\ref{eq: scaling form}) 
and (\ref{eq: C_red_inf_decay}), the scaling behaviour of the reduced 
correlation function is
\begin{equation}
C^{\text{(red)}}_N(l) = \sqrt{ C_\infty(l) } f(l) + \frac{g(l)}{N} + \ldots
\label{eq: C_red_scaling}
\end{equation}
It follows that the reduced separation
$\left\langle l \right\rangle_N^{\text{(red)}}$ scales 
linearly with $N$. This is clearly shown in Fig.\
\ref{Fig: red ph sep scaling}, where $\left\langle l \right\rangle_N^{\text{(red)}}$ is plotted 
for the $1B_u$ and $2A_g$ states.


\begin{references}
\bibitem{foot_dipole}
The dipole operator is defined as: 
$\hat{\mu}=\sum_{i} x_i \hat{n}_{i}$, where $x_i$ is the coordinate
of the $i$th site along the polymer axis and $\hat{n}_{i}$ is the 
occupation number operator.
\bibitem{foot_averaging}
The averaging is performed in order to take into account the two 
different odd bond lengths in the dimerised lattice.
\bibitem{foot_ionicity}
Our definition differs from that of \cite{McWilliams} by a factor of 
$N/2$.
\bibitem{foot_minimum}
The minimum separation is one chemical bond.
\bibitem{foot_excitation}
This is true for any excitation which does not involve the promotion of 
a finite fraction the particles from the Fermi sea across the gap.
\end{references}
\end{document}